\documentclass[%
aip,
twocolumn,
amsmath,amssymb,
groupedaddress,
 reprint,%
]{revtex4-2}

\usepackage{graphicx}
\usepackage{bm}
\usepackage{color}
\usepackage{extarrows}
\usepackage{mathptmx}

\renewcommand{\d}{{\rm d}}

\newcommand{\w}{\omega}

\newcommand{\ti}{\tilde}

\newcommand{\la}{\langle}
\newcommand{\ra}{\rangle}
\newcommand{\lla}{\langle\langle}
\newcommand{\rra}{\rangle\rangle}

\newcommand{\tL}{\textrm{L}}
\newcommand{\tR}{\textrm{R}}
\newcommand{\tT}{\textrm{T}}
\newcommand{\tB}{\textrm{B}}

\newcommand{\tSB}{\textrm{SB}}
\newcommand{\tS}{\textrm{S}}
\newcommand{\tC}{\textrm{C}}
\newcommand{\B}{\textrm{B}}
\newcommand{\A}{\textrm{A}}
\newcommand{\T}{\textrm{T}}
\newcommand{\Tr}{\textrm{Tr}}
\newcommand{\SB}{\textrm{SB}}
\newcommand{\rmeq}{\textrm{eq}}
\newcommand{\st}{\textrm{st}}

\newcommand{\tr}{\mathrm{tr}}

\newcommand{\nl}{\nonumber \\}
\newcommand{\eq}[1]{Eq.\,(\ref{#1})}
\newcommand{\Eq}[1]{Equation \,(\ref{#1})}

\newcommand{\Fig}[1]{Fig.\,\ref{#1}}

\newcommand{\RN}[1]{
  \textup{\uppercase\expandafter{\romannumeral#1}}
}

\begin{document}
\author{Zi-Hao Chen}
\affiliation{Department of Chemical Physics University of Science and Technology of China, Hefei, Anhui 230026, China}
\author{YiJing Yan}
\affiliation{Department of Chemical Physics University of Science and Technology of China, Hefei, Anhui 230026, China}
\email{yanyj@ustc.edu.cn}
\title
{
	Kondo regime of the impurity spectral function and the current noise spectrum in the double impurity Anderson model
}
\date{\today}

\begin{abstract}
	The dissipaton equations of motion (DEOM) method is one of the most popular methods for simulating quantum impurity systems.
	In this article, we use DOEM theory to deal with the Kondo problem of the double quantum dots (DQDs) impurity system.
	We focus on the impurity spectral function and the total noise spectral function, this two function will be used to describe the Kondo effect of this system.
	The influence of the interaction, the hooping and the difference of the chemical potential between the two dots on the Kondo effect of the system is studied.
	We find that the interaction between the two dots can influence the Kondo effect of the system a lot.
\end{abstract}
\maketitle

\section{Introduction}
An efficient impurity solver is highly required in strong correlation problems. A variety of numerical simulation methods can be applied as impurity solvers \cite{Wil75773, Kri801003, Kri801044, Bul08395, Hir862521, Gul11349, Whi922863, Han19050601, Wan012979, Muh08176403, Ema11349}.
Among them, the most famous approaches are the numerical renormalization group (NRG) \cite{Wil75773, Kri801003, Kri801044}, the density matrix renormalization group (DMRG) \cite{Whi922863} and quantum Monte Carlo (QMC) \cite{Ema11349}.
This method can obtain single--particle Green function and two--particle correlation function efficiency.
But the dynamic properties, such as dynamic I-V characteristics differential conductance, and propagation of density matrix, can not be obtained directly using NRG or QMC. And as it is very time-consuming to obtain the two--particle correlation function such as the current noise spectrum.

The real--time propagation approaches basically make up for this shortcoming.
One can directly obtain multi-particle correlation functions, such as the current noise spectrum, which cannot be obtained by NRG or QMC.
It includes the hierarchical equations of motion (HEOM) \cite{Tan906676, Tan06082001, Yan04216, Xu05041103, Xu07031107, Jin08234703} and its second quantization version dissipaton equation of motion (DEOM) \cite{Yan16110306}, the semi-group quantum master equations\cite{Lin76119, Gor76821,Ali87}, and the DMFT methods \cite{Hou14045141}. Although the DMRG and NRG have their time--dependent extension, these methods still have shortcomings in efficiency compared to HEOM/DEOM \cite{Xu22230601}.

The strong correlation problems contain varieties of systems, among them, quantum dots have been widely studied not only in computations but also in experiments.\cite{Wie031, Han071217, Rei021283}
As an interesting and popular system, single quantum dots and multiple quantum dots (MQDs) can be regarded as ``artificial atoms'' and ``artificial molecules''.\cite{Bli967899, Wie031, Jeo012221}
The QDs and reservoirs around them can form strongly correlated quantum systems. Especially,  those interactions lead to the famous Kondo resonance at low temperatures.\cite{Lia02725,Far20256805, Moc21186804, Kur216004, Fer20738}
Depending on the specific system, QDs can form {Ruderman--Kittel--Kasuya--Yosida} indirect exchange interactions,\cite{Pow1349} inter--dot Coulomb interactions and intra--dot Coulomb interactions (capacitive interactions).\cite{Hew93}
Those two types of coupling strongly influence the properties of systems.

In this article, we utilize the DEOM method, one of the most popular methods for simulating quantum impurity systems, to simulate the density of state (DOS) and the noise spectrum of double quantum dots.

\section{Dissipaton equations of motion}
The Hamiltonian of the open quantum system can be written as
\begin{equation}
	H_{\T} = H_{\tS} + H_{\SB} + H_{\tB},
\end{equation}
where $H_{\tS}$ is the Hamiltonian of the system, $H_{\tB}$ is the Hamiltonian of the bath, and $H_{\SB}$ is the Hamiltonian of the coupling between the system and the bath.
In the Anderson impurity model (AIM), $H_{\tS}$ can be arbitrary, $H_{\SB}$ and $H_{\tB}$ can be written as
\begin{subequations}
	\begin{align}
		h_{\tB}           & = \sum_{\alpha} h_{\alpha} = \sum_{\alpha k} \epsilon_{\alpha k} \hat d_{\alpha k}^{+} \hat d_{\alpha k},                                                                        \\
		H_{\tSB}          & = \sum_{\alpha u} (\hat F^{\dagger}_{\alpha u} \hat a_{u} + \hat a^{\dagger}_{u} \hat F_{\alpha u}) = \sum_{\sigma \alpha u} \hat a_{u}^{\bar \sigma} \ti F^{\sigma}_{\alpha u}, \\
		\hat F_{\alpha u} & = \sum_{k} t_{\alpha u k}^{\ast} \hat d_{\alpha k},
	\end{align}
\end{subequations}
where $\hat a_{u}$ ($\hat a_{u}^{\dagger}$) is the annihilation (creation) operator of the system electron.
Note $u$ labels the degrees of freedom of the system electrons, which can be the spin or the site index.
The fluctuation--dissipation theorem of this bath can be written as
\begin{equation}
	\label{heom:fermi_dft}
	\la \hat F_{\alpha u}^{\sigma}(t) \hat F_{\alpha v}^{\bar \sigma}(0) \ra_{\tB}^{\textrm{eq}} = \frac{1}{\pi} \int \d \w \frac{J_{\alpha u v}^{\sigma}(\w) e^{i \sigma \w t}}{1 + e^{\sigma \beta \w}},
\end{equation}
where, $\beta = K_{B} T$, $K_{B}$ is the Boltzmann constant, $\sigma = \pm 1$ is the fermion sign and $J_{\alpha u v}^{\sigma}(\w)$ is the spectral density of the bath, $F_{\alpha u}^{\sigma}(t) = e^{ih_{\tB}t} F_{\alpha u} e^{-ih_{\tB}t}$ and $\la \hat O \ra_{\tB}^{\rmeq} = \tr_{\tB} (\hat O e^{-\beta_{\alpha} {\hat h}_{\alpha}}) / Z_{\alpha}^{\rmeq}$ with the canonical ensembles partition function $Z_{\alpha}^{\rmeq} = \tr_{\tB} e^{-\beta_{\alpha} {\hat h}_{\alpha}}$.

In the DEOM theory, we expand this time correlation function as the summation of exponentials:
\begin{equation}
	\label{eq:c_t}
	\la \hat F_{\alpha u}^{\sigma}(t) \hat F_{\alpha v}^{\bar \sigma}(0) \ra_{\tB}^{\textrm{eq}} = \sum_{k=1}^{K} \eta^{\sigma}_{\alpha u v k} e^{- \gamma^{\sigma}_{\alpha u v k} t} = \sum_{j = 1}^{J} n_{j} \gamma_{j}.
\end{equation}
Then, the DEOM formulism reads as follows:\cite{Yan16110306}
\begin{align}
	\label{eq:DEOM}
	\dot \rho_{\bf n}^{(n)} (t) = & (- i \mathcal{L}_{\tS} - \sum_{j}n_{j}\gamma_{j})\rho^{(n)}_{\bf n} - i \sum_j \mathcal{A}_{\bar j} \rho^{(n+1)}_{{\bf n}j} \nl
	                              & - i \sum_{j} (-1)^{n-\theta_j} \mathcal{C}_{j} \rho^{(n-1)}_{{\bf n}_j^-}.
\end{align}
Throughout this paper, we set $\hbar = 1$.
In the above equation, the summation of j from 1 to J, where J is exactly the number of terms in \eq{eq:c_t}. The $\{\rho_{\bf n}^{(n)}\}$ are the dissipaton density operators (DDOs). $\theta_j = \sum_{k = 1} ^{j} n_{k}$. $\mathcal{L}_{\tS} \hat O = [H_{\tS}, \hat O]$ and the other super operators are defined as:
\begin{subequations}
	\label{eq:super_operator}
	\begin{align}
		\mathcal{A}_{j} \rho^{(n)}_{\bf n} & \equiv\hat a_u^{\sigma} \rho^{(n)}_{\bf n} + (-1)^n\rho^{(n)}_{\bf n} \hat a_u^{\sigma},                                                                                      \\
		\mathcal{C}_{j} \rho^{(n)}_{\bf n} & \equiv \sum_v \big( \eta^{\sigma}_{\alpha u k} \hat a_u^{\sigma} \rho^{(n)}_{\bf n} - (-1)^n \eta^{\bar \sigma \ast}_{\alpha u k}  \rho^{(n)}_{\bf n} \hat a_u^{\sigma}\big),
	\end{align}
\end{subequations}
where $\hat a_{u}^{\dagger}$ ($\hat a_{u}$) is the local creation (annihilation) operator of the $i\ $th system electron with spin $s$, here $s= \,\uparrow$ and $\downarrow$.
Note the $(-1)^n$ factor in the definition of super operators in \eq{eq:super_operator} is due to the fermion sign.

The DEOM theory can also deal with the nonequilibrium steady-state case. The nonequilibrium bath can be described by the following eﬀective Hamiltonian:
\begin{equation}
	h^{\st}_{\tB} = \sum_{\alpha} h^{\st}_{\alpha} = \sum_{\alpha k} (\epsilon_{\alpha k} + \mu_{\alpha}) \hat d_{\alpha k}^{+} \hat d_{\alpha k}.
\end{equation}
Then \eq{heom:fermi_dft} can be recast as follows:
\begin{equation}
	\label{heom:fermi_dft_neq}
	\la \hat F_{\alpha u}^{\sigma}(t) \hat F_{\alpha v}^{\bar \sigma}(0) \ra^{\st}_{\tB} = \frac{1}{\pi} \int \d \w \frac{J_{\alpha u v}^{\sigma}(\w - \mu_{\alpha}) e^{i \sigma \w t}}{1 + e^{\sigma \beta (\w - \mu_{\alpha})}}.
\end{equation}
Note in the above equation, $F_{\alpha u}^{\sigma}(t) = e^{ih^{\st}_{\tB}t} F_{\alpha u} e^{-ih^{\st}_{\tB}t}$ and $\la \hat O \ra_{\tB}^{\st} = \tr_{\tB} (\hat O e^{-\beta_{\alpha} {\hat h}^{\st}_{\alpha}}) / Z_{\alpha}^{\st}$ with the grand canonical ensembles partition function $Z_{\alpha}^{\st} = \tr_{\tB} e^{-\beta_{\alpha} {\hat h}^{\st}_{\alpha}}$.
We can obtain
\begin{equation}
	\la \hat F_{\alpha u}^{\sigma}(t) \hat F_{\alpha v}^{\bar \sigma}(0) \ra^{\st}_{\tB} = e^{i \sigma \mu_{\alpha} t} \la \hat F_{\alpha u}^{\sigma}(t) \hat F_{\alpha v}^{\bar \sigma}(0) \ra^{\rmeq}_{\tB},
\end{equation}
and the other relationships and the equation of motion keep the same as the equilibrium case.

In the numerical simulation of the DEOM method, we need to truncate the DDOs to a finite level $L$, which means that we only keep the DDOs, $\rho_{\bf n}^{(n)}$, satisfying $\sum_k n_k < L$ and discard all the other DDOs.
Thus, in the numerical simulation of the fermionic bath, the number of DDOs is
\begin{equation}
	\sum_{l=1}^{L} \frac{K!}{l! (k-l)!}.
\end{equation}
Here the $K$ is the bath modes (See \eq{eq:c_t}) and the $L$ is the truncation level.
The vast of DDOs makes the direct simulation of the DEOM method very expensive.
But with the matrix product state (MPS) and time-dependent variational principle (TDVP) methods \cite{Ose112295, Lub15917, Shi18174102, Xu22230601}, the cost of propagating DEOM can be reduced to nearly proportional to $K$.
Although the MPS method will be the slower one under the small $K$ cases.
In this article, we will only show the results of simulations using the direct method.
We use the recently developed time-domain Prony Fitting Decomposition ($t$-PFD) method to obtain the parameters of the summation of exponentials. \cite{Che22221102}
$t$-PFD method can obtain the almost minimal basis of the summation of exponential.

In this article, we focus on the impurity spectral function,
\begin{equation}
	A_{u u'}(\w) = \frac{1}{2 \pi} \int_{-\infty}^{\infty}\!\!\d t\, e^{i\w t} \la\{ \hat a_{u} (t), \hat a_{u'}^{\dagger} (0)\} \ra_{\rmeq},
\end{equation}
and the noise spectral function
\begin{equation}
	\label{eq::noise_spectral_function}
	S_{\alpha \alpha'}(\w) = \frac{1}{2 \pi} \int_{-\infty}^{\infty}\!\!\d t\, e^{i\w t} \la\{\delta \hat I_{\alpha} (t), \delta \hat I_{\alpha'} (0)\} \ra_{\rmeq}.
\end{equation}
Here $\la \hat O \ra_{\rmeq} = \Tr (\hat O \rho_{\T}^{\rmeq})$. In \eq{eq::noise_spectral_function}, $\delta \hat I_{\alpha} (t) = \hat I_{\alpha} (t) - \hat I_{\alpha}^{\textrm{st}}$ is the fluctuation of the transport current respect to the steady state current $\hat I_{\alpha}^{\textrm{st}}$.
The transport current is defined as
\begin{equation}
	\hat I_{\alpha} = - \frac{\partial \hat N_\alpha}{\partial t} = - i \sum_{u} (\hat a_{u}^{\dagger} \hat F_{\alpha u} - \hat F_{\alpha u}^{\dagger} \hat a_{u})
\end{equation}
and $\hat I_{\alpha}(t) = e^{i H_{\textrm{\T}}t} \hat I_{\alpha} e^{-i H_{\T}t}$.

Both the impurity and the noise spectral functions can be calculated using the DEOM method. The details of those correlation functions by the DEOM method can be found in \onlinecite{Jin15234108, Yan16110306, Mao21014104}. In this article, we use the self--consistent iteration method \cite{Zha17044105} to obtain the equilibrium state and the spectral density function. See Appendix for details.

\section{Impurity and noise spectral function of double quantum dots}

\begin{figure}[t]
	\centering
	\includegraphics[width = 0.48\textwidth]{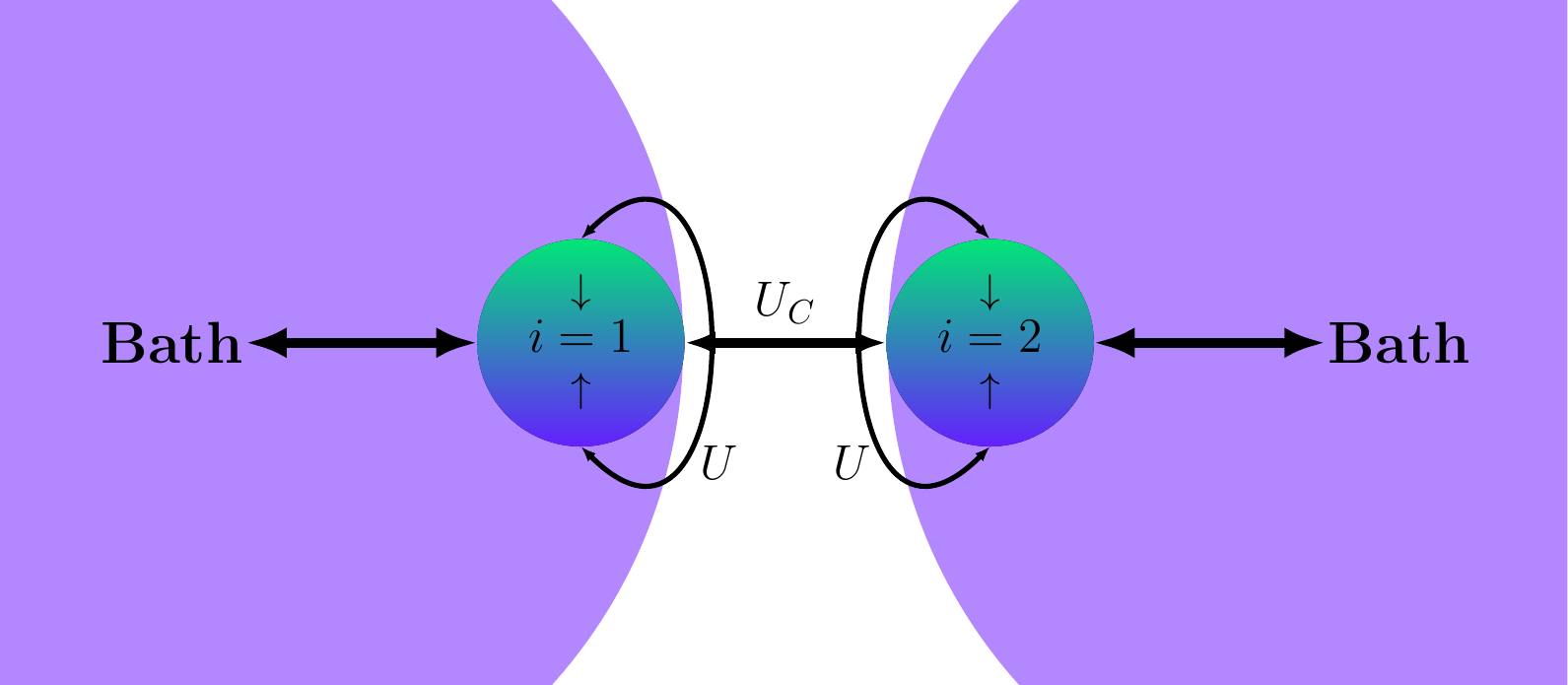}
	\caption{
		The illustration of the AIM with the system as double quantum dots.}
	\label{fig1}
\end{figure}

As a numerical demonstration, we choose the DQD for simulations. To be concrete, we set
\begin{align} \label{sys_def}
	H_{\tS} = & \sum_{i = 1,2} \epsilon_i \hat n_i + U \sum_{i = 1,2} \hat n_{i\uparrow} \hat n_{i\downarrow} + U_{\tC} \hat n_{1} \hat n_{2} \nl
	          & + T \sum_{s} \Big(\hat a^{\dagger}_{1s} \hat a_{2s} + \hat a^{\dagger}_{2s} \hat a_{1s} \Big),
\end{align}
where $\hat n_{i,\uparrow}$ ($\hat n_{i\downarrow}$) is the particle number operator of spin-up (spin-down) electron at site $i$ and $\hat n_{i} = \hat n_{i\uparrow} + \hat n_{i\downarrow}$
$\epsilon_i$ is on--site energy, $U$ ($U_{\tC}$) is intra--site (inter--site) Coulomb energy, and T is the hopping energy between two sites.
The system parameters are follow this scheme: $\epsilon_{1} = \epsilon_{2} = -(U + 2NU_{\tC})/2$.
This leads to $\la \hat n_{1} + \hat n_{2} \ra = N + 1$ in the equilibrium states of the system.

The system is coupled to two reservoirs with the Lorentz--type spectrum density
\begin{equation}
	\label{eq:lorentz}
	J_{\alpha u v}^{\sigma}(\w) = \frac{\Delta W^2}{\w^2 + W^2}.
\end{equation}
We set the left and right reservoirs link to the 1st quantum dot and the 2ed quantum dot, respectively.
For convenience, we label the left and right reservoirs as $1$ and $2$, respectively.
These two bath parameters are the same: $W = 50 \Delta$, $U = 12 \Delta$, $\beta = 20 \Delta^{-1}$ with $\Delta$ as the unit.
The total current noise spectrum which follows the Ramo--Shockley theorem \cite{Bla001} can be written as
\begin{align}
	\label{eq::noise}
	S(\w) & = a^2 S_{\tL \tL}(\w) + b^2 S_{\tR \tR}(\w) - 2ab \mathrm{Re} \{S_{\tL \tR}(\w)\},
\end{align}
The parameter $a$ and $b$ are relative to the left and right leads and the impurity, respectively.
In the wideband limit, the coupling strength of the two reservoirs is the same leading to $a = b = 0.5$ in \eq{eq::noise}.

The total system is illustrated in Fig.~\ref{fig1}.
We set the truncated level as $L = 5$, and the number of exponential series as $K = 6$, which guarantee the error between the summation of the exponential expansion and the original one less than $2\%$ under all frequency.

In this section, we show both the impurity and noise spectral function of double quantum dots under equilibrium scenarios with or without hopping and nonequilibrium scenarios.
We will show that the impurity spectra will split

\subsection{Equilibrium Scenarioes}

\begin{figure}[t]
	\centering
	\includegraphics{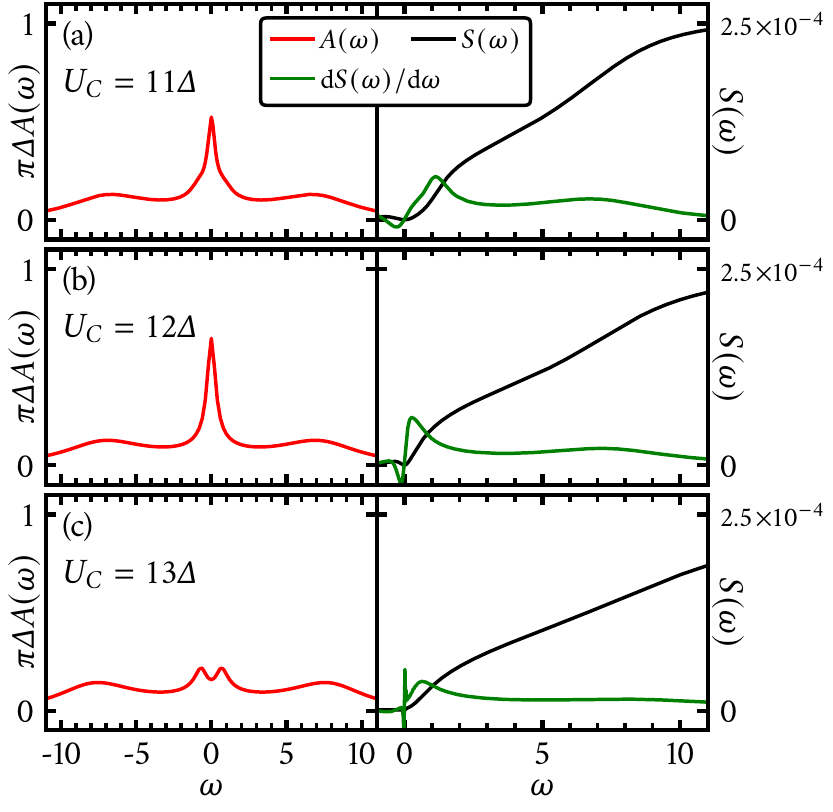}
	\caption{
		The impurity spectral function $A(\w) = A_{1 \uparrow, 1 \uparrow}(\w)$ (left panels), the total noise spectral function $S(\w)$ (middle panels), and the derivation of the total noise spectral function (right panels) of the AIM with the system as double quantum dots.
		The bath parameters as $W = 50 \Delta$, $U = 12 \Delta$, $U_{\tC} = 11 \Delta$, $12 \Delta$, $13 \Delta$, $N = 1$, $T_{\tC} = 0$ and $\beta = 20 \Delta^{-1}$.
		The system parameters are follow this scheme: $\epsilon_{1} = \epsilon_{2} = -(U + 2NU_{\tC})/2$.
		The other parameters are shown in each panel.
		Here, we set the truncated level $L = 5$.}
	\label{fig2}
\end{figure}

\begin{figure*}
	\centering
	\includegraphics{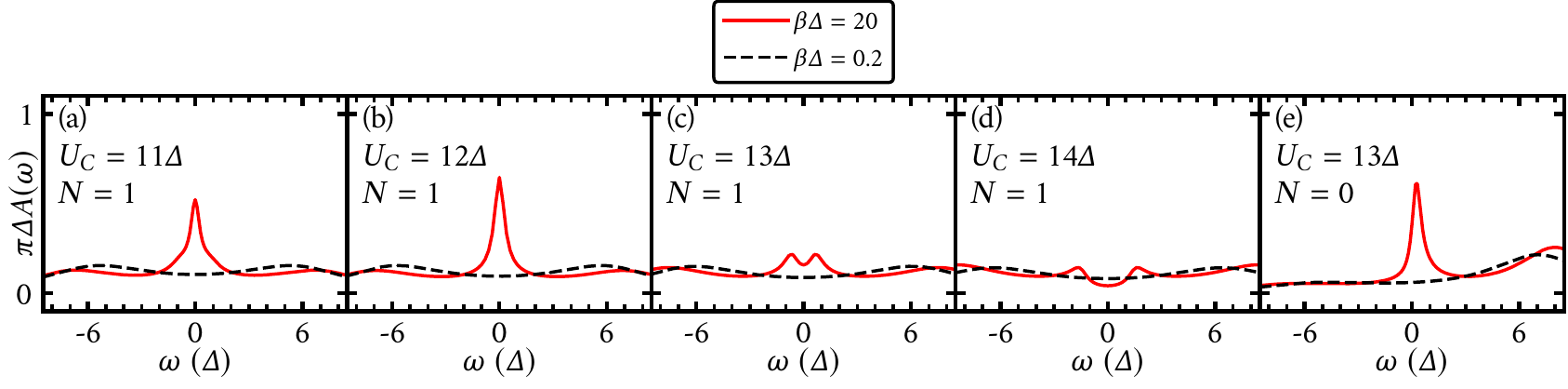}
	\caption{
		The impurity spectral function $A(\w) = A_{1 \uparrow, 1 \uparrow}(\w)$. We set the parameters as $U = 12 \Delta$, $U_{\tC} = 11 \Delta$, $12 \Delta$, $13 \Delta$, $14 \Delta$ under $N = 1$ scenario, $U_{\tC} = 11 \Delta$ under $N = 0$ scenario and $\beta = 0.2 \Delta^{-1}$, $20 \Delta^{-1}$.
		We set the other parameters are the same as those in the \Fig{fig2}.
	}
	\label{fig2-2}
\end{figure*}

In the \Fig{fig2}, we show the impurity spectral function $A_{1 \uparrow, 1 \uparrow}(\w)$ (left panels), the total noise spectral function $S(\w)$ (middle panels) of the AIM with the system as double quantum dots.
We also show the derivation of the total noise spectral function, $\d S(\w) / \d t$, at the right panel at \Fig{fig2}.
Here, we set $U_{\tC} = U - \Delta$, $U$, $U + \Delta$, and the chemical potential of these two baths as both $0$ (equilibrium scenario).
We also check the results of $A(\w)$ using the MPS method (not shown in the \Fig{fig2}) and we can obtain similar results as those shown in the \Fig{fig2}.

We can see that with the $U_{\tC}$ increase, the Kondo peak, $A(0)$, will first increase and then decrease.
The highest peak takes place at $U_{\tC} = U$, which shows the resonance effect of the double quantum dots.
The Hubbard peak, which is the peak at $\w \approx \pm U/2$, will move far from $0$.
We also observed the new Kondo peak, appearing around $\w \approx \pm (U - U_{\tC})$ at $U_{\tC} = U \pm \Delta$.
This peak is the result of the inter--site Coulomb interaction $U_{\tC}$ and will disappear when $U_{\tC} = U$.
The behaviors of the Kondo peak near the $U_{\tC} = U$ are similar to the Fano resonance.
When $U_{\tC} > U$, the electron transfer between the two quantum dots will be blocked by the inter--site Coulomb interaction $U_{\tC}$, which leads to the heavily decrease of the Kondo peak.

To illustrate these phenomena, we compare the results of the impurity spectral function under more settings at different temperatures in \Fig{fig2-2}.
We can see that all the Kondo peaks will disappear at low temperatures.
This is because the Kondo effect is a low--temperature phenomenon and shows those split peaks in \Fig{fig2} are not Hubbard peaks.
Under the $N = 0$ scenario, those split peaks will vanish and the behavior of the Kondo peak will be similar to \Fig{fig2}\,(b).
This behavior is due to the absence of the Coulomb blockade under the $N = 0$ scenario.
The equilibrium state of those double quantum dots is $\la \hat n_{1} \ra = \la \hat n_{2} \ra = (N + 1) / 2$.
In the $N = 1$ scenario, $\la \hat n_{1} \ra = \la \hat n_{2} \ra = 1$.
The symmetry of the impurity spectral function, $A(\w) = A(-\w)$, is also broken; See the panel (e) of \Fig{fig2-2}.

Now turn to the total noise spectral function $S(\w)$, which is shown in the right panels of \Fig{fig2}.
The Kondo characteristics of the total noise spectral function will show at the derivation of $S(\w)$, $\d S(\w) / \d t$, the total noise spectral function will behave like the step function, and the Kondo peak or the Hubbard peaks will appear in the derivation of $S(\w)$. \cite{Jin15234108}
As shown in the \Fig{fig2}, the Hubbard peaks occur near $\w \approx U/2$, and will decrease with the increase of $U_{\tC}$.
The Kondo peak will appear near $\w \approx 0$.
The location of the Kondo peak is influenced by the $U_{\tC}$. 
Under the $U_{\tC} = U$ scenario, the Kondo peak will appear exactly at $\w = 0$.
Under the $U_{\tC} = U - \Delta$ case, the Kondo peak will move to $\w \approx \Delta$, and remain a small Kondo peak at $\w = 0$.
The $U_{\tC} = U + \Delta$ case is similar to the $U_{\tC} = U - \Delta$ case, but under this case, the Kondo peak near $\w = 0$ becomes very large.

\subsection{Equilibrium Scenarioes with Hopping}

\begin{figure}[b]
	\centering
	\includegraphics{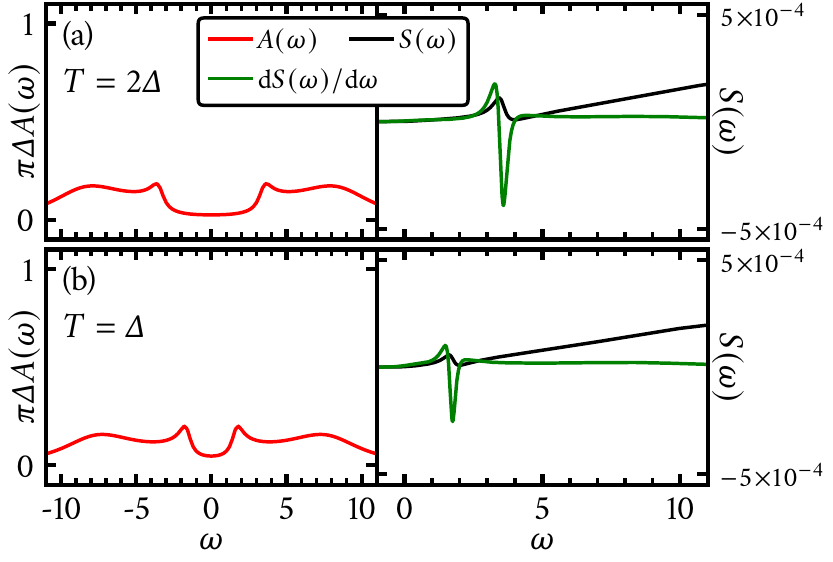}
	\caption{
		The impurity spectral function $A(\w) = A_{1 \uparrow, 1 \uparrow}(\w)$ (left panels), the total noise spectral function $S(\w)$ (right panels), and the derivation of the total noise spectral function (right panels) of the AIM with the system as double quantum dots.
		We set the hopping energy as $T_{\tC} = 0.5 \Delta$ and $\Delta$, $U_{\tC} = U$, and the other parameters are the same as those in the \Fig{fig2}.
	}
	\label{fig3}
\end{figure}

In the \Fig{fig3}, we show the impurity spectral function $A_{1 \uparrow, 1 \uparrow}(\w)$ (left panels) and the total noise spectral function $S(\w)$ (right panels) of the AIM with the system as double quantum dots.
Here, we set the hopping energy as $T_{\tC} = 0.5 \Delta$ or $\Delta$, $U_{\tC} = U$, and the other parameters are the same as those in the \Fig{fig2}.
We also notice that the sign of $T_{\tC}$ will not influent the result of both $A(\w)$ and $S(\w)$.

As shown in the \Fig{fig3}, the Kondo peak of the impurity spectral function will split into two peaks, which appear at $\w \approx \pm 2 T = \pm 2 \Delta$, $\pm 4 \Delta$.
These phenomena are similar to what we show in \Fig{fig2}(c). But under this scenario, these behaviors are due to the inducement of $T_{\tC}$ causing an antiferromagnetic interaction and then the splitting of the Kondo peak. \cite{Li18115133}
The Hubbard peaks remain occur at a similar location as $U_{\tC} = U$ case in the \Fig{fig2}\,(b).

Turning to the total noise spectral function $S(\w)$, we only focus on the derivation of this type of spectral function, which is shown in the right panels of \Fig{fig3}.
The derivation of the total noise spectral function, $\d S(\w) / \d t$, will also split into two peaks, which appear at $\w \approx 0$ and $\w \approx 2 T = \Delta$.
The Hubbard peaks is absent around $\w \approx U/2$.
Moreover, the total noise spectral function will perform like Fano resonance other than the step function.

\begin{figure}[t]
	\centering
	\includegraphics{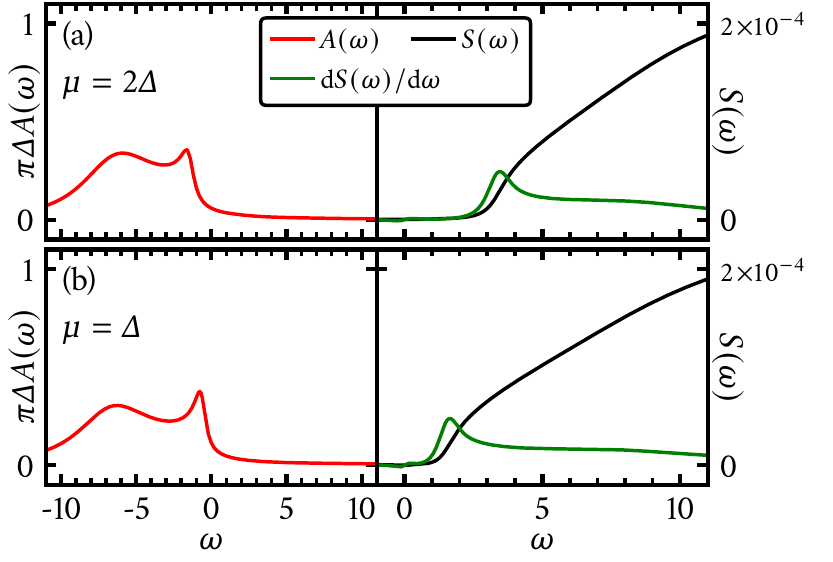}
	\caption{
		The impurity spectral function $A(\w) = A_{1 \uparrow, 1 \uparrow}(\w)$ (left panels), the total noise spectral function $S(\w)$ (right panels), and the derivation of the total noise spectral function (right panels) of the AIM with the system as double quantum dots.
		We set the bias as $\mu_{1} = -\mu_{2} = \Delta$ and $2 \Delta$, $T_{\tC} = 0$, $U_{\tC} = U$, and the other parameters are the same as those in the \Fig{fig2}.
	}
	\label{fig4}
\end{figure}

\subsection{Nonequilibrium Scenarioes}

In the \Fig{fig4}, we show the impurity spectral function $A_{1 \uparrow, 1 \uparrow}(\w)$ (left panels) and the total noise spectral function $S(\w)$ (right panels) of the AIM with the system as double quantum dots.
We notice that $A_{2 \uparrow, 2 \uparrow}(\w) = A_{1 \uparrow, 1 \uparrow}(- \w)$ (not shown in the \Fig{fig4})
Here, we set the chemical potential of these two baths as $\mu_{L} = - \mu_{R} = \Delta$ and $2 \Delta$ with the other parameters same as \Fig{fig2}.

As shown in the \Fig{fig4}, the Kondo peak of the impurity spectral function will move to $\w \approx \pm \Delta$ and $\pm 2 \Delta$.
This is the same behavior as the single quantum dot case,\cite{Wan13035129} but with only the $\w < 0$ part, and the  $\w > 0$ part will appear in the $A_{2 \uparrow, 2 \uparrow}(\w)$.
Turning to the derivation of current noise spectral function, which is shown in the right panels of \Fig{fig4}.
The Kondo peak will move to $\w \approx 2 \mu$ and will keep a little bit of the Fano resonance behavior near $\w = 0$.
The Hubbard peaks is absent around $\w \approx U/2$.

\section{Summary}
In this article, we use the nearly developed time domain Prony fitting decomposition method and the self--consistent tteration method to simulate the Anderson impurity model with double quantum dots.

Support from the Ministry of Science and Technology of China (Grant No.\ 2021YFA1200103) and the National Natural Science Foundation of China (Grant Nos. 22103073, 22173088) is gratefully acknowledged.
The numerical calculations in this paper have been done on the supercomputing system in the Supercomputing Center of University of Science and Technology of China.

\appendix*
\section{Self--consistent Iteration Method}
The self--consistent iteration method has been used to solve the equilibrium state of the bosonic environmental DEOM.\cite{Zha17044105}
We will utilize this method to solve both the equilibrium state and the spectral density function of the fermionic environmental DEOM.

Firstly, we show the workflow for solving the equilibrium state of the fermionic environmental DEOM.
The equilibrium state of the fermionic environmental DEOM is defined as
\begin{align}
	\label{heom:steady_state_example_1}
	0 =
	 & (- i \mathcal{L}_{\tS} - \sum_{j}n_{j}\gamma_{j})\rho^{(n)}_{\bf n} - i \sum_j \mathcal{A}_{\bar j} \rho^{(n+1)}_{{\bf n}j} \nl
	 & - i \sum_{j} (-1)^{n-\theta_j} \mathcal{C}_{j} \rho^{(n-1)}_{{\bf n}_j^-}.
\end{align}
This equation can be rewritten as
\begin{align}
	\label{heom:steady_state_example_2}
	\Big(i \mathcal{L}_{\tS} + \gamma_{\bf n} + \Omega\Big)\rho^{(n)}_{\bf n}
	= & \Omega \rho^{(n)}_{\bf n} - i \sum_j \mathcal{A} \rho^{(n+1)}_{{\bf n}_j^+} \nl
	  & - i \sum_{j} n_{j} \mathcal{C}_{j} \rho^{(n-1)}_{{\bf n}_j^-}.
\end{align}
We add the stability factor $\Omega$ to both sides of the above equation.
We can use the iterative method to solve the above equation.
\begin{align}
	\label{heom:steady_state_example_3}
	\rho^{(n); i+1}_{\bf n} = & \Big(i \mathcal{L}_{\tS} + \gamma_{\bf n} + \Omega\Big)^{-1}\Big(\Omega \rho^{(n); i }_{\bf n} - i \sum_j \mathcal{A} \rho^{(n+1); i}_{{\bf n}_j^+} \nl
	                          & - i \sum_{j} n_{j} \mathcal{C}_{j} \rho^{(n-1); i}_{{\bf n}_j^-}\Big)
\end{align}
The stability factor, $\Omega$, can make this iterative method stable.

Secondly, We can use the same method to solve the spectral density function.
The spectral density function is defined as
\begin{align}
	\label{heom:corr_freq}
	\hat C_{\A \B} (\w)
	 & \equiv \frac{1}{\pi} \int_{0}^{\infty} \Tr \Big\{{\hat A} e^{- i \mathcal{L}_{T} t} {\hat B} \rho_{\tT}^{\rmeq} \Big\} e^{i \w t} \d t\nl
	 & = \frac{1}{\pi} \int_{0}^{\infty} \Tr \Big\{{\hat A} e^{i (\w - \mathcal{L}_{T}) t} {\hat B} \rho_{\tT}^{\rmeq} \Big\} \d t \nl
	 & = \frac{1}{\pi} \Tr \Big\{{\hat A} (i \mathcal{L}_{T} - i \w)^{-1} {\hat B} \rho_{\tT}^{\rmeq} \Big\} \nl
	 & = \frac{1}{\pi} \lla \hat {\bm A} \vert \hat {\bm X}(\w) \rra,
\end{align}
where,
\begin{align}
	\label{heom:ddo_operator}
	\rho_{\tT}(t) & \rightarrow {\bm \rho}(t) \equiv \{\rho_{\bf n}^{(n)}(t)\}, \nl
	\hat A        & \rightarrow \hat {\bm A} \equiv \{\hat A^{(n)}_{\bf n}; n = 0,1,2,\cdots\}, \nl
	\hat B        & \rightarrow \hat {\bm B} \equiv \{\hat B^{(n)}_{\bf n}; n = 0,1,2,\cdots\},
\end{align}
and
\begin{align}
	\label{heom:spe_w_sch}
	(i {\bm{\mathcal{L}}}_{T} - i \w) \hat {\bm X}(\w) = {\hat B} \rho^{\rmeq}_{\tT} = {\bm{\rho}}(0; \hat B).
\end{align}
\Eq{heom:spe_w_sch} can be rewritten as follows:
\begin{align}
	\label{heom:sci_w_example}
	\rho^{(n)}_{\bf n} (0; \hat B)
	= & - (i \mathcal{L}_{\tS} + \gamma_{\bf n} + i \w) \hat X^{(n)}_{\bf n}(\w) \nl
	  & - i \sum_j \mathcal{A}_{j} \hat X^{(n+1)}_{{\bf n}_j^+} (\w) \nl
	  & - i \sum_{j} (-1)^{n-\theta_j} \mathcal{C}_{j} \hat X^{(n-1)}_{{\bf n}_j^-}(\w).
\end{align}
as the same procedure as (\ref{heom:steady_state_example_1}) --\eq{heom:steady_state_example_3}, we can obtain the similar iterative form:
\begin{align}
	\label{heom:spe_w_sch_final}
	\hat X^{(n); i+1}_{\bf n}(\w)
	= & (i \mathcal{L}_{\tS} + \gamma_{\bf n} + i \w + \Omega)^{-1} \Bigg\{- \rho^{(n)}_{\bf n} (0; \hat B) \nl
	  & + \Omega \hat X^{(n); i}_{\bf n}(\w)  - i \sum_j \mathcal{A}_{j} \hat X^{(n+1); i}_{{\bf n}_j^+}(\w) \nl
	  & - i \sum_{j} (-1)^{n-\theta_j} \mathcal{C}_{j} \hat X^{(n-1); i}_{{\bf n}_j^-}(\w) \Bigg\}.
\end{align}


\begin{thebibliography}{10}

	\bibitem{Wil75773}
	K.~G. Wilson, \newblock ``The renormalization group: Critical phenomena and
	  Kondo problem,'' Rev. Mod. Phys. {\bf 47}, 773 (1975).
	
	\bibitem{Kri801003}
	H.~R. Krishna-murthy, J.~W. Wilkins, and K.~G. Wilson, \newblock
	  ``Renormalization-group approach to the Anderson model of dilute magnetic
	  alloys. I. Static properties for the symmetric case,'' Phys. Rev. B {\bf 21},
	  1003 (1980).
	
	\bibitem{Kri801044}
	H.~R. Krishna-murthy, J.~W. Wilkins, and K.~G. Wilson, \newblock
	  ``Renormalization-group approach to the Anderson model of dilute magnetic
	  alloys. II. Static properties for the asymmetric case,'' Phys. Rev. B {\bf
	  21}, 1044 (1980).
	
	\bibitem{Bul08395}
	R.~Bulla, T.~A. Costi, and T.~Pruschke, \newblock ``Numerical renormalization
	  group method for quantum impurity systems,'' Rev. Mod. Phys. {\bf 80}, 395
	  (2008).
	
	\bibitem{Hir862521}
	J.~E. Hirsch and R.~M. Fye, \newblock ``Monte Carlo method for magnetic
	  impurities in metals,'' Phys. Rev. Lett. {\bf 56}, 2521 (1986).
	
	\bibitem{Gul11349}
	E.~Gull, A.~J. Millis, A.~I. Lichtenstein, A.~N. Rubtsov, M.~Troyer, and
	  P.~Werner, \newblock ``Continuous-time Monte~Carlo methods for quantum
	  impurity models,'' Rev. Mod. Phys. {\bf 83}, 349 (2011).
	
	\bibitem{Whi922863}
	S.~R. White, \newblock ``Density matrix formulation for quantum renormalization
	  groups,'' Phys. Rev. Lett. {\bf 69}, 2863 (1992).
	
	\bibitem{Han19050601}
	L.~Han, V.~Chernyak, Y.~A. Yan, X.~Zheng, and Y.~J. Yan, \newblock ``Stochastic
	  representation of non-Markovian fermionic quantum dissipation,'' Phys. Rev.
	  Lett. {\bf 123}, 050601 (2019).
	
	\bibitem{Wan012979}
	H.~Wang, M.~Thoss, and W.~H. Miller, \newblock ``Systematic convergence in the
	  dynamical hybrid approach for complex systems: A numerically exact
	  methodology,'' J. Chem. Phys. {\bf 115}, 2979 (2001).
	
	\bibitem{Muh08176403}
	L.~M{\"u}hlbacher and E.~Rabani, \newblock ``Real-time path integral approach
	  to nonequilibrium many-body quantum systems,'' Phys. Rev. Lett. {\bf 100},
	  176403 (2008).
	
	\bibitem{Ema11349}
	E.~Gull, A.~J. Millis, A.~I. Lichtenstein, A.~N. Rubtsov, M.~Troyer, and
	  P.~Werner, \newblock ``Continuous-time Monte Carlo methods for quantum
	  impurity models,'' Rev. Mod. Phys. {\bf 83}, 349 (2011).
	
	\bibitem{Tan906676}
	Y.~Tanimura, \newblock ``Nonperturbative expansion method for a quantum system
	  coupled to a harmonic-oscillator bath,'' Phys. Rev. A {\bf 41}, 6676 (1990).
	
	\bibitem{Tan06082001}
	Y.~Tanimura, \newblock ``Stochastic Liouville, Langevin, Fokker-Planck, and
	  master equation approaches to quantum dissipative systems,'' J. Phys. Soc.
	  Jpn. {\bf 75}, 082001 (2006).
	
	\bibitem{Yan04216}
	Y.~A. Yan, F.~Yang, Y.~Liu, and J.~S. Shao, \newblock ``Hierarchical approach
	  based on stochastic decoupling to dissipative systems,'' Chem. Phys. Lett.
	  {\bf 395}, 216 (2004).
	
	\bibitem{Xu05041103}
	R.~X. Xu, P.~Cui, X.~Q. Li, Y.~Mo, and Y.~J. Yan, \newblock ``Exact quantum
	  master equation via the calculus on path integrals,'' J. Chem. Phys. {\bf
	  122}, 041103 (2005).
	
	\bibitem{Xu07031107}
	R.~X. Xu and Y.~J. Yan, \newblock ``Dynamics of quantum dissipation systems
	  interacting with bosonic canonical bath: Hierarchical equations of motion
	  approach,'' Phys. Rev. E {\bf 75}, 031107 (2007).
	
	\bibitem{Jin08234703}
	J.~S. Jin, X.~Zheng, and Y.~J. Yan, \newblock ``Exact dynamics of dissipative
	  electronic systems and quantum transport: Hierarchical equations of motion
	  approach,'' J. Chem. Phys. {\bf 128}, 234703 (2008).
	
	\bibitem{Yan16110306}
	Y.~J. Yan, J.~S. Jin, R.~X. Xu, and X.~Zheng, \newblock ``Dissipaton equation
	  of motion approach to open quantum systems,'' Frontiers Phys. {\bf 11},
	  110306 (2016).
	
	\bibitem{Lin76119}
	G.~Lindblad, \newblock ``On the generators of quantum dynamical semigroups,''
	  Commun. Math. Phys. {\bf 48}, 119 (1976).
	
	\bibitem{Gor76821}
	V.~Gorini, A.~Kossakowski, and E.~C.~G. Sudarshan, \newblock ``Completely
	  positive dynamical semigroups of $N$-level systems,'' J. Math. Phys. {\bf
	  17}, 821 (1976).
	
	\bibitem{Ali87}
	R.~Alicki and K.~Lendi,
	\newblock {\em Quantum Dynamical Semigroups and Applications: Lecture Notes in
	  Physics 286},
	\newblock Springer, New York, 1987.
	
	\bibitem{Hou14045141}
	D.~Hou, R.~L. Wang, X.~Zheng, N.~H. Tong, J.~H. Wei, and Y.~J. Yan, \newblock
	  ``Hierarchical equations of motion for impurity solver in dynamical
	  mean-field theory,'' Phys. Rev. B {\bf 90}, 045141 (2014).
	
	\bibitem{Xu22230601}
	M.~Xu, Y.~Yan, Q.~Shi, J.~Ankerhold, and J.~T. Stockburger, \newblock ``Taming
	  {{Quantum Noise}} for {{Efficient Low Temperature Simulations}} of {{Open
	  Quantum Systems}},'' Phys. Rev. Lett. {\bf 129}, 230601 (2022).
	
	\bibitem{Wie031}
	W.~G. van~der Wiel, S.~D. Franceschi, J.~M. Elzerman, T.~Fujisawa, S.~Tarucha,
	  and L.~P. Kouwenhoven, \newblock ``Electron transport through double quantum
	  dots,'' Rev. Mod. Phys. {\bf 75}, 1 (2003).
	
	\bibitem{Han071217}
	R.~Hanson, L.~P. Kouwenhoven, J.~R. Petta, S.~Tarucha, and L.~M.~K.
	  Vandersypen, \newblock ``Spins in few-electron quantum dots,'' Rev. Mod.
	  Phys. {\bf 79}, 1217 (2007).
	
	\bibitem{Rei021283}
	S.~M. Reimann and M.~Manninen, \newblock ``Electronic structure of quantum
	  dots,'' Rev. Mod. Phys. {\bf 74}, 1283 (2002).
	
	\bibitem{Bli967899}
	R.~H. Blick, R.~J. Haug, J.~Weis, D.~Pfannkuche, K.~V. Klitzing, and K.~Eberl,
	  \newblock ``Single-electron tunneling through a double quantum dot: The
	  artificial molecule,'' Phys. Rev. B {\bf 53}, 7899 (1996).
	
	\bibitem{Jeo012221}
	H.~Jeong, A.~M. Chang, and M.~R. Melloch, \newblock ``The Kondo effect in an
	  artificial quantum dot molecule,'' Science {\bf 293}, 2221 (2001).
	
	\bibitem{Lia02725}
	W.~Liang, M.~P. Shores, M.~Bockrath, J.~R. Long, and H.~Park, \newblock ``Kondo
	  resonance in a single-molecule transistor,'' Nature {\bf 417}, 725 (2002).
	
	\bibitem{Far20256805}
	L.~Farinacci, G.~Ahmadi, M.~Ruby, G.~Reecht, B.~W. Heinrich, C.~Czekelius,
	  F.~von Oppen, and K.~J. Franke, \newblock ``Interfering tunneling paths
	  through magnetic molecules on superconductors: Asymmetries of Kondo and
	  Yu-Shiba-Rusinov resonances,'' Phys. Rev. Lett. {\bf 125}, 256805 (2020).
	
	\bibitem{Moc21186804}
	C.~P. Moca, I.~Weymann, M.~A. Werner, and G.~Zar\'and, \newblock ``Kondo cloud
	  in a superconductor,'' Phys. Rev. Lett. {\bf 127}, 186804 (2021).
	
	\bibitem{Kur216004}
	A.~Kurzmann, Y.~Kleeorin, C.~Tong, R.~Garreis, A.~Knothe, M.~Eich, C.~Mittag,
	  C.~Gold, F.~K. {de Vries}, K.~Watanabe, T.~Taniguchi, V.~Falko, Y.~Meir,
	  T.~Ihn, and K.~Ensslin, \newblock ``Kondo effect and spin--orbit coupling in
	  graphene quantum dots,'' Nat. Comm. {\bf 12}, 1 (2021).
	
	\bibitem{Fer20738}
	M.~Ferrier, R.~Delagrange, J.~Basset, H.~Bouchiat, T.~Arakawa, T.~Hata,
	  R.~Fujiwara, Y.~Teratani, R.~Sakano, A.~Oguri, K.~Kobayashi, and R.~Deblock,
	  \newblock ``Quantum noise in carbon nanotubes as a probe of correlations in
	  the Kondo regime,'' J. Low Temp. Phys. {\bf 201}, 738 (2020).
	
	\bibitem{Pow1349}
	S.~R. Power and M.~S. Ferreira, \newblock ``Indirect exchange and
	  Ruderman--Kittel--Kasuya--Yosida (RKKY) interactions in magnetically-doped
	  graphene,'' Crystals {\bf 3}, 49 (2013).
	
	\bibitem{Hew93}
	A.~C. Hewson,
	\newblock {\em The Kondo Problem to Heavy Fermions},
	\newblock Cambridge University Press, Cambridge, 1993.
	
	\bibitem{Ose112295}
	I.~V. Oseledets, \newblock ``Tensor-{{Train Decomposition}},'' SIAM J. Sci.
	  Comput. {\bf 33}, 2295 (2011).
	
	\bibitem{Lub15917}
	C.~Lubich, I.~V. Oseledets, and B.~Vandereycken, \newblock ``Time
	  {{Integration}} of {{Tensor Trains}},'' SIAM J. Numer. Anal. {\bf 53}, 917
	  (2015).
	
	\bibitem{Shi18174102}
	Q.~Shi, Y.~Xu, Y.~Yan, and M.~Xu, \newblock ``Efficient Propagation of the
	  Hierarchical Equations of Motion Using the Matrix Product State Method,'' The
	  Journal of Chemical Physics {\bf 148}, 174102 (2018).
	
	\bibitem{Che22221102}
	Z.-H. Chen, Y.~Wang, X.~Zheng, R.-X. Xu, and Y.~Yan, \newblock ``Universal
	  Time-Domain {{Prony}} Fitting Decomposition for Optimized Hierarchical
	  Quantum Master Equations,'' J. Chem. Phys. {\bf 156}, 221102 (2022).
	
	\bibitem{Jin15234108}
	J.~S. Jin, S.~K. Wang, X.~Zheng, and Y.~J. Yan, \newblock ``Current noise
	  spectra and mechanisms with dissipaton equation of motion theory,'' J. Chem.
	  Phys. {\bf 142}, 234108 (2015).
	
	\bibitem{Mao21014104}
	H.~Mao, J.~Jin, S.~Wang, and Y.~Yan, \newblock ``Nonequilibrium {{Kondo}}
	  Regime Current Noise Spectrum of Quantum Dot Systems with the Single Impurity
	  {{Anderson}} Model,'' J. Chem. Phys. {\bf 155}, 014104 (2021).
	
	\bibitem{Zha17044105}
	H.~D. Zhang, Q.~Qiao, R.~X. Xu, X.~Zheng, and Y.~J. Yan, \newblock ``Efficient
	  steady-state solver for hierarchical quantum master equations,'' J. Chem.
	  Phys. {\bf 147}, 044105 (2017).
	
	\bibitem{Bla001}
	{\relax Ya.M}.~Blanter and M.~B{\"u}ttiker, \newblock ``Shot Noise in
	  Mesoscopic Conductors,'' Physics Reports {\bf 336}, 1 (2000).
	
	\bibitem{Li18115133}
	Z.~Li, Y.~Cheng, J.~Wei, X.~Zheng, and Y.~Yan, \newblock ``Kondo-Peak Splitting
	  and Resonance Enhancement Caused by Interdot Tunneling in Coupled Double
	  Quantum Dots,'' Phys. Rev. B {\bf 98}, 115133 (2018).
	
	\bibitem{Wan13035129}
	S.~K. Wang, X.~Zheng, J.~S. Jin, and Y.~J. Yan, \newblock ``Hierarchical
	  Liouville-space approach to nonequilibrium dynamic properties of quantum
	  impurity systems,'' Phys. Rev. B {\bf 88}, 035129 (2013).
	
	\end{thebibliography}
\end{document}